\documentclass[useAMS,usenatbib]{mn2e}

\usepackage{amsmath,amssymb}
\usepackage{amsfonts}
\usepackage{bbold}
\usepackage{graphicx}
\usepackage{color}
\usepackage{hyperref}
\usepackage{transparent}
\usepackage{rotating}
\usepackage{caption}
\usepackage{subcaption}

\usepackage{xspace}
\newcommand*{\eg}{e.g.\@\xspace}
\newcommand*{\ie}{i.e.\@\xspace}

\title[2.5D B-H accretion onto a compact object]{Numerical simulations of axisymmetric hydrodynamical Bondi-Hoyle accretion onto a compact object}
\author[I. El Mellah and F. Casse]{I. El Mellah$^{1}$\thanks{\href{mailto:ileyk@apc.univ-paris7.fr}{ileyk@apc.univ-paris7.fr}} \& F. Casse$^{1}$\\
$^{1}$Laboratoire AstroParticule et Cosmologie - Universit\'e Paris 7 Diderot - 10 rue Alice Domon et L\'eonie Duquet, Paris, 75013, France}
\begin{document}

\date{Accepted 2015 September 16.  Received 2015 September 16; in original form 2015 August 13}

\pagerange{\pageref{firstpage}--\pageref{lastpage}} \pubyear{YEAR}

\maketitle

\label{firstpage}

\begin{abstract}
Bondi-Hoyle accretion configurations occur as soon as a gravitating body is immersed in an ambient medium with a supersonic relative velocity. From wind-accreting X-ray binaries to runaway neutron stars, such a regime has been witnessed many times and is believed to account for shock formation, the properties of which can be only marginally derived analytically. In this paper, we present the first results of the numerical characterization of the stationary flow structure of Bondi-Hoyle accretion onto a compact object, from the large scale accretion radius down to the vicinity of the compact body. For different Mach numbers, we study the associated bow shock. It turns out that those simulations confirm the analytical prediction by \cite{Foglizzo1996} concerning the topology of the inner sonic surface with an adiabatic index of 5/3. They also enable us to derive the related mass accretion rates, the position and the temperature of the bow shock, as function of the flow parameters, along with the transverse density and temperature profiles in the wake.
\end{abstract}


\section{Introduction}

High energy astronomy has turned the spotlight on accretion processes and in particular, on the interaction of compact objects with their surroundings. Their capacity of gravitationally gathering the ambient gas made those bodies indirectly visible through the light emitted by the heated accreted material. The study of this interplay aims at broadening our understanding of those exotic objects as much as probing the fuelling sources. When such an interlacing takes place, the simplest forms to consider are systems where a homogeneous flow of matter is submitted to the gravitational field of an isolated point-mass in an inertial frame. If the Bondi's spherical model of a non-zero temperature stationary flow at infinity helps to clarify the situation \citep{Bondi1952}, the relative motion of the accreting body compared to its environment should be accounted for, which led Hoyle, Lyttleton and afterwards, Bondi to lay the foundations of an axisymmetric model of a zero temperature supersonic flow at infinity (henceforward referred to as {\sc b-h} or wind accretion) in publications staggered over 5 years starting in the late 30's \citep{Hoyle:1939fl,Bondi1944} ; thermodynamics was sacrificed in favour of a more realistic symmetry. As X-ray astronomy was born and the existence of compact objects was firmly established, the predictions of their model turn out to be accurate enough to drive the community into refining it and assess the origin of possible discrepancies (inhomogeneities, pressure effects, orbital considerations, etc) : the wind accreting systems family grew.

The harvest of X-ray data provided by the last decade satellites not only spotted previously unseen sources but also promoted some accreting systems which were, up to now, merely marginal cases, to the rank of representatives of a well-established observational family. It is, \eg, the case for the super-giant X-ray binaries ({\sc s}g{\sc xb}), a kind of wind accreting systems which number has increased by almost a factor of three thanks to {\sc integral} and \textit{Swift} monitoring \citep{Walter15}. The last few years have also seen an increasing interest into Hyper-Luminous X-ray sources ({\sc hlx}) which seem to spectrally behave in a qualitatively similar way as galactic X-ray binaries hosting black hole candidates, exhibiting low/hard and high/soft states \citep{VanderMarel2003,MillerC2004,Webb2014}. The specificities of the accretion flows in those systems remains an open question and some of them might be runaway Intermediate-Mass Black Holes ({\sc imbh}) undergoing wind accretion. While runaway neutron stars have already been observed in the Milky Way \citep{Cordes1993}, galactic runaway {\sc imbh} are still to be found. It was also suggested by \cite{Pfahl2002} that accretion of a stellar companion wind by neutron stars could be the main culprit for the low-luminosity hard X-rays point sources detected with \textit{Chandra} by \cite{Wang2002} in the Galactic Center region and studied in greater details more recently \citep{Degenaar2012,Degenaar2015}.

In parallel, the new computational capacities offer us the opportunity to unveil the physical phenomena at stake behind the scenes with the search for numerical setups able to reproduce the observational classification. For the last decades, {\sc b-h} accretion has been extensively studied both with analytical and numerical approaches. The latter started to blossom in the late 80's with a series of works investigating the influence of the hydrodynamical ({\sc hd}) terms. The stability of the {\sc b-h} flow was put into question with numerical results glimpsing various instabilities without being able to agree whether their origin was physical or not \citep[see][and references therein]{Foglizzo2005}. Meanwhile, analytical progresses were also made as physical effects were added or coupled together : small or large scale magnetic fields \citep{Igumenshchev2002,Igumenshchev2006,Pang2011}, net vorticity \citep{Krumholz2005}, radiative feedback \citep{Park2013}, finite size accretors \citep{Ruffert1994a}, turbulence \citep{Krumholz2006}, etc. Simulations of accretion onto not compact objects like stars orbiting an {\sc agb} companion in a symbiotic binary \citep{Theuns1996,deValBorro:2009gk} have deeply improved our understanding of the barium enriched stars \citep{Bidelman1951} ; but the wide dynamics introduced by the small size of a compact object compared to its gravitational sphere of influence on non-relativistic winds has prevented numerical simulations to converge to a {\sc b-h} accretion solution around a compact object up to now. A major theoretical step was undertaken with the study of spherical accretion of a flow with a non vanishing velocity at infinity, mixing ballistic and pressure effects all together \citep{Theuns1992}. The same blending remained to be done for axisymmetric geometries until \cite{Foglizzo1996}, henceforth FR96, linked the sonic surface\footnote{To be understood as the surface where the flow speed oversteps the local sound speed.} of a {\sc b-h} flow to the simpler one of the spherical configuration. Section \ref{sec:theory} will be devoted to an introduction to this topological result. 

In this paper, our primary goal is to design a robust axisymmetric numerical setup taking the most of the high performance computing methods implemented in the {\sc mpi}-parallelized Adaptive Mesh Refinement Versatile Advection Code ({\sc mpi}-{\sc amrvac}) presented in Section \ref{sec:methods}. We constantly monitored our results to compare them to firmly established analytical constraints, derived accurate numerical mass accretion rates and studied the steady state structure of the flow (transverse profiles, sonic surface, etc). Using a continuously nested logarithmic mesh with a shock-capturing algorithm and specific boundary conditions, we can resolve the flow from the vicinity of the central compact object up to the significant deflection scale, at an affordable computational price and with realistic wind velocities. In agreement with the properties of {\sc b-h} accretion on finite size objects recalled in \cite{Ruffert1994a}, we can finally reach numerical regimes where the size of the inner boundary no longer significantly alters the flow properties. We solve the full set of {\sc hd} Eulerian equations, including the energy one, with the less restrictive adiabatic assumption instead of considering a polytropic flow and identifying the adiabatic and the polytropic indexes. We investigate the evolution of the properties of the {\sc b-h} accretion flows we obtain with the Mach number of the flow at infinity, from slightly subsonic to highly supersonic setups, and summarize our results in Section \ref{sec:results}.    

\section[]{Theoretical framework}
\label{sec:theory}
\subsection{Hoyle-Lyttleton accretion}
\begin{figure}
\begin{center}
\includegraphics[width=0.25\textwidth]{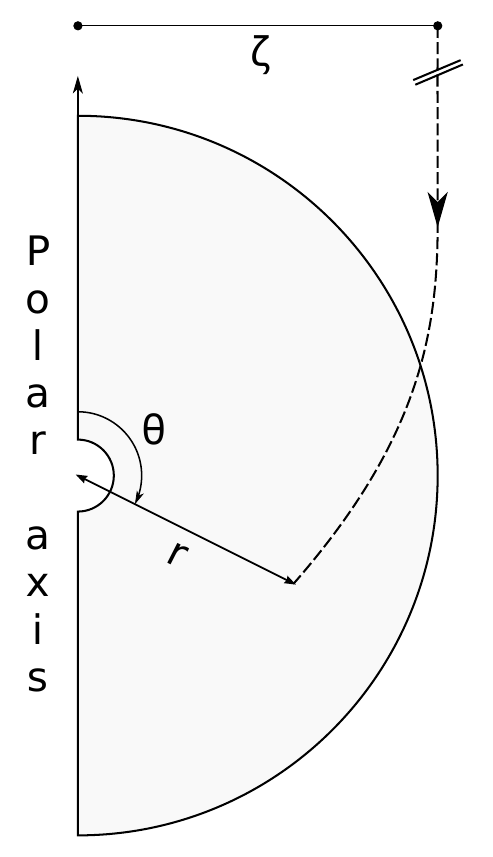}
\caption{Sketch illustrating the ballistic {\sc b-h} flow with the variables we consider. A fiducial streamline corresponding to the ballistic velocities \eqref{eq:vr} and \eqref{eq:vt} for a given impact parameter $\zeta$ is plotted in dashed. The light grey area is the simulation space. For the sake of visibility, its inner boundary is highly oversized compared to its outer one.}
\label{fig:sketch}
\end{center}
\end{figure}

Back in 1939, Hoyle \& Lyttleton portrayed the streaming of a homogeneous supersonic planar flow with a relative velocity and a density at infinity of $\mathbf{v_{\infty}}$ and $\rho _{\infty}$, deflected by the gravitational field of a point mass $M$. As a first approximation, they neglected the influence of the {\sc hd} terms and solved the equations of motion for a test-mass \citep{Hoyle:1939fl}. Such an axisymmetric ballistic approach leads to the explicit velocity field and trajectory, as further described by the Figure\,\ref{fig:sketch} and in the section \ref{sec:BC} \citep{Bisnovatyi-Kogan1979}. Yet, once the flow reaches the line lying downstream of the gravitational body, dissipative effects are likely to lead to a substantial damping of the orthoradial component of the velocity field, letting the test-mass worn out with a total mechanical energy which turns out to be negative for particles with an impact parameter $\zeta$ verifying the Hoyle-Lyttleton accretion condition :
\begin{equation}
\label{eq:racc_HL}
\zeta<\zeta_{\textsc{hl}}=\frac{2GM}{v_{\infty}^2}
\end{equation}
$\zeta_{\textsc{hl}}$ is commonly referred to as the accretion radius or the stagnation point even if, in this picture, particles are left with a non-zero velocity provided by the remaining radial component. $G$ stands for the usual gravitational constant. Given the symmetry of the problem, one can expect that all independent particles in a cylinder of cross section $\pi \zeta_{\textsc{hl}}^2$ will eventually be trapped in the gravitational potential. Those considerations led Hoyle \& Lyttleton to suggest an accretion rate of :
\begin{equation}
\label{eq:Mdot_HL}
\dot{M}_{\textsc{hl}}=\pi \zeta_{\textsc{hl}}^2 \rho _{\infty} v_{\infty}= \frac{4\pi G^2 M^2 \rho_{\infty}}{v_{\infty}^3}
\end{equation}
Due to the severe ballistic assumption, such a model does not describe the motion of the particles once they reached the infinitely thin and dense accretion line in the wake of the point mass, when thermodynamical effects (pressure and dissipation) prevail. 


\subsection{Accretion column model}
The stationary isotropic accretion of a zero velocity but non-zero temperature flow at infinity was first studied by \cite{Bondi1952} who determined the corresponding accretion rate \citep[see][for a derivation]{Frank2002} :

\begin{equation}
\label{eq:MBONDI}
\dot{M_{\textsc{b}}}=\frac{4\pi G^2 M^2 \rho_{\infty}}{c_{\infty}^3} \left( \frac{1}{2} \right)^{\frac{\gamma +1}{2\left( \gamma -1 \right)}} \left( \frac{4}{5-3\gamma} \right)^{\frac{5-3\gamma }{2\left( \gamma -1 \right)}}
\end{equation}
where $c_{\infty}$ is the sound speed at infinity, uniquely determined by the adiabatic index $\gamma$ of the flow and the temperature at infinity\footnote{Via, for an ideal gas : $c^2:=\frac{\partial P}{\partial \rho}\bigg|_{S}=\gamma \frac{P}{\rho} \propto T$}. The similarities between the mass accretion rates formulae \eqref{eq:MBONDI} and \eqref{eq:Mdot_HL} led Bondi to suggest a first interpolation formula between those respectively spherical and axisymmetric asymptotic cases, as a first empirical attempt to account for thermal and kinetic effects all together, latter refined by \cite{Shima1985} :
\begin{equation}
\label{eq:interpolBH}
\dot{M}_{\textsc{bh}}=\dot{M}_{\textsc{hl}} \left( \frac{\mathcal{M_{\infty}}^2}{\mathcal{M_{\infty}}^2+1} \right)^\frac{3}{2}  
\end{equation}
where $\mathcal{M_{\infty}}$ is the Mach number of the incoming flow, $v_{\infty}/c_{\infty}$.This formula matches the Hoyle-Lyttleton mass accretion rate at high Mach numbers but not the Bondi one for low Mach numbers. 

A more analytical approach was undertaken in the context of the accretion column models which detail the role of pressure forces on the dynamics of a non-zero thickness wake \citep{Bondi1944,Wolfson1977,Yabushita1978,Yabushita1978a,Horedt2000,Edgar:2005bi}. Indeed, the compression of mildly supersonic flows is bound to lead to the formation of a shock in the vicinity of $\theta\sim\pi$, especially as the Mach number is large, where the pressure gradient is likely to play a significant role. Those two-flow models (supersonic Hoyle-Lyttleton deflected flow supplying matter and momentum / one-dimensional channelled flow along the column) lead to a non linear system of ordinary differential equations describing the dynamics of the accretion column ; investigations of the latter suggests that the position of the stagnation point and thereby, the value of the mass accretion rate, could vary by an order-of-unity factor around the Hoyle Lyttleton mass accretion rate given by \eqref{eq:Mdot_HL}. Yet, for the sake of solvability, those studies either neglect the pressure force in the wake \citep{Edgar:2005bi} or assume a polytropic relation between pressure and density to bypass the energy equation \citep{Horedt2000}. 

\subsection{Sonic surface of B-H flows}
In the spherical Bondi accretion formalism, the sonic radius, which can be obtained, \eg, from the equation (2.11b) of \cite{Theuns1992}, locates the uniquely determined position where a stationary subsonic inflowing gas becomes supersonic :
\begin{equation}
\label{eq:r0}
r_0=\frac{5-3\gamma}{4}\frac{GM}{\displaystyle c_{\infty}^2+\frac{\gamma -1}{2}v_{\infty}^2}
\end{equation}
The intrinsic multi-dimensionality of the {\sc b-h} flow has prohibited until now any comparable level of detail, concerning the analytical properties of the sonic surface, to be reached. FR96 started to bridge the analytical gap between the spherical Bondi accretion of a flow at rest at infinity and the ballistic pressureless Hoyle-Lyttleton accretion, as they proved a topological property of the sonic surface of {\sc b-h} accretion flows in relation with the radius of their isotropic counterpart, $r_0$, with the same Bernoulli's invariant value\footnote{Notice that the Bernoulli's invariant is the same for all streamlines due to homogeneous conditions at infinity.}. They showed that the {\sc b-h} sonic surface must intersect at least once the sphere of radius $r_0$. The main analytical constraint entailed by this property is the necessity for the sonic surface of flows with $\gamma=5/3$ to be anchored to the inner boundary of a simulation, whatever its size since the sonic radius vanishes. 
FR96 also derived an interpolation formula based on asymptotic expansions near the accretor and considerations about the sonic surface ; its expression is reminded in the section \ref{sec:mdot} where we compare the mass accretion rates from the simulations to the ones predicted by this interpolation formula. FR96 suggest, providing an unknown constant of unity-order, that the interpolation formula they determined could fit the stationary mass accretion rates measured.

\subsection{Dealing with the energy}
For the sake of simplicity, most papers until now have reported on results based on the polytropic assumption\footnote{Or an isothermal one.} to relate the pressure to the density ; if this swindle in analytical studies is an almost compulsory assumption to go forwards in the computation, it is not as mandatory when it comes to numerical simulations. The only requirement is to add an additional equation to solve without adding any complexity (provided the flow is assumed to be adiabatic - see comments below), which might already unveil results beyond the theoretical expectations. Indeed, the main drawback of the polytropic assumption is to add an additional degree of freedom with the polytropic index $\Gamma$. $\Gamma$ has no reason to be equal to the adiabatic index $\gamma$ (determined by the chemical nature of the flow), as long as the flow is not isentropic \citep{Horedt2000}, which is not our case here because of the expected shock formation ; $\Gamma$ does not even need to be homogeneous in space since in different places will dominate different cooling/heating processes. Less restrictive than the equality between $\gamma$ and $\Gamma$ is the adiabatic assumption : entropy will not be exchanged within the flow but can still appear at shock fronts, where the variables are discontinuous enough to introduce irreversibility. If this hypothesis alleviates the introduction of an additional degree of freedom and enables us to naturally handle the shock jump conditions, the price to pay is to restrain the comparison of our results to prediction relying on $\gamma=5/3$ for it is the adiabatic index of any flow hot enough to be composed of monoatomic (if not, ionized) particles. Besides, our $\gamma=5/3$ simulations may still depart from theoretical descriptions since we homogeneously describe the flow via an energy equation while writing $P\propto\rho ^{\gamma}$ precludes any conclusive evidence on the effects of the shock. In terms of phenomenological limitations, the method we chose also limits the results to compact objects undergoing \textit{adiabatic} wind accretion. 

\section[]{Designing axisymmetric simulations of BH accretion}
\label{sec:methods}
\subsection{MPI-AMRVAC}
\label{sec:vac}
We derive the structure of the {\sc b-h} accretion flow relying on the {\sc mpi-amrvac} code, further described in \cite{Porth:2014wv}. Briefly, the {\sc mpi}-{\sc amrvac} package consists in a multi-dimensional finite-volume code able to solve, in multiple geometries, the set of equations of hydrodynamics or magnetohydrodynamics for an ideal gas, either in a classical or relativistic framework.
In the present paper, we called upon a shock-capturing second order in time and space Total Vanishing Diminishing Lax-Friedrichs ({\sc tvdlf}) scheme and a Koren slope limiter \citep{Koren1993} to solve the previously mentioned conservative {\sc hd} equations in a robust and accurate way : 

\begin{equation}
\label{eq:eq1}
\partial _t \rho + \boldsymbol{\nabla} \cdot \left( \rho \mathbf{v} \right) = 0
\end{equation}
\begin{equation}
\label{eq:eq2}
\partial _t \left( \rho \mathbf{v} \right) + \boldsymbol {\nabla} \cdot \left( \rho \mathbf{v} \otimes \mathbf{v} + P \mathbb{1} \right) = - \rho \boldsymbol {\nabla} \Phi
\end{equation}
\begin{equation}
\label{eq:eq3}
\partial _t  e  + \boldsymbol{\nabla} \cdot \left[ \left( e + P \right) \mathbf{v} \right] = - \rho \mathbf{v} \cdot \boldsymbol {\nabla} \Phi
\end{equation}
with :
\begin{enumerate}
  \item[-]  the mass density $\rho$
  \item[-] the velocity $\mathbf{v}$
  \item[-] the total specific energy $e=e_K + u$ with $e_K$ the kinetic energy per volume unit and $u$ being the internal energy per volume unit related to the pressure $P$ via the ideal gas equation-of-state
  \item[-] the gravitational potential of the accreting object of mass $M$, $\Phi(r)=-\frac{GM}{r}$ (no self-gravity)
\end{enumerate}
As explained in the previous section, the energy equation we rely on does not contain any heat transfer term ; it is adiabatic. Computing them would slow down the simulations but, more importantly, would require either a full radiative transfer equation solver or strong assumptions about the predominant cooling/heating terms. We chose to bypass the latter which is likely to require refinement as soon as the internal energy density variations due to heat exchange are no longer negligible compared to the ones due to the work done by pressure forces. 
\subsection{Numerical setup and boundary conditions}
\subsubsection{Axisymmetric spherical ($r$,$\theta$) coordinates}
\label{sec:2.5D}
Since the current paper deals with the axisymmetric properties of the {\sc b-h} accretion flow, we can work on a spherical 2.5D mesh, where the third component is the longitudinal one, $\varphi$. Thus, the cells do respect the full 3D geometry information (in the expression of the curved interfaces and the implied divergence terms in \eqref{eq:eq1} - \eqref{eq:eq3} for instance) but the flow is assumed to be invariant by rotation around the polar axis, which spares the mesh to span a third dimension. The incoming wind velocity at infinity is then collinear to the polar axis and chosen to move towards the South (\ie comes from $\theta \to 0$ towards $\theta \to \pi$). Given the reported reasonable stability of the 3D {\sc b-h} flow against transverse instabilities as the "flip-flop" one at stake in some 2D cylindrical {\sc b-h} configuration \citep{Blondin2009}, both theoretically \citep{Soker1990} and numerically \citep{Blondin:2012vf}, we expect such a configuration to give birth to a relatively robust equilibrium from which full 3D simulation should not depart much.

The initial flow and the outer radial boundary are attributed a zero longitudinal velocity, $v_{\varphi}$. The corresponding evolution equation \eqref{eq:eq2}$\cdot\boldsymbol{\hat{\varphi}}$ guarantees that the analytical $\rho v_{\varphi}$ remains null, which is confirmed in all the following simulations ; this variable, yet computed, will not be discussed any further. Regarding the other conservative quantities, they evolve accordingly to the simplified 2.5D spherical form of the equation in \eqref{eq:eq1} - \eqref{eq:eq3}.
\subsubsection{Grid resolution}
A regular equidistant 2.5D spherical grid would suffer two major drawbacks. First, to get a decent radial resolution next to the inner boundary, we would have to work with a huge number of radial cells, which would be painfully time-consuming and would not tell us much more about the relevant dynamics. In addition, the cell aspect ratio $\Delta r / r \Delta \theta$ cannot remain constant all along a radius ; $r$ varying from the inner to the outer boundary by a factor $10^2$ to $10^4$ (depending on the simulation), a regular grid would give highly deformed cells. As indicated by its name, {\sc mpi-amrvac} presents the capacity to adaptively refine the mesh where needed but it would introduce spurious discontinuities in the grid resolution.

A safer alternative \citep[initially introduced for {\sc b-h} flows by][]{Fryxell1988} is to work with a constant aspect ratio, in other words, with a radial step proportional to the radius (hence the "logarithmic grid" designation). It enables us to keep a homogeneous "relative resolution", from the vicinity of the compact object up to the accretion radius. We typically work with $N_{\theta}=64$ latitudinal cells and $N_r=128$, $176$ or $224$ radial cells, depending on the size of the inner radius of the grid ; once the latter is set, we tune the aspect ratio around the unity value to make the outer radius approximately constant from a simulation to another. Given the angular resolution, this approach leads to a inner radial cell sizes of the order of a twentieth of the inner boundary radius.
\subsubsection{Boundary conditions}
\label{sec:BC}
The partial differential equations from \eqref{eq:eq1}-\eqref{eq:eq3} are associated with usual polar symmetric and antisymmetric conditions in $\theta=0$ and $\pi$, using the frame specified on Figure\,\ref{fig:sketch}. For the outer radial boundary conditions, far upstream, where the flow is supersonic, we prescribe the ballistic solution for $v_r$ and $v_{\theta}$ and the permanent regime solution deduced for $\rho$ from the mass conservation equation by \cite{Bisnovatyi-Kogan1979} :
\begin{equation}
\label{eq:vr}
v_r=-v_{\infty}\sqrt{1+\frac{\zeta_{\textsc{hl}}}{r}-\left( \frac{\zeta}{r} \right)^2 }
\end{equation}
\begin{equation}
\label{eq:vt}
v_{\theta}=v_{\infty}\frac{\zeta}{r}
\end{equation}
\begin{equation}
\label{eq:dens_BK}
\rho=\rho_{\infty} \frac{\zeta ^2}{r\sin\theta \left( 2\zeta-r\sin\theta \right)}
\end{equation}
with $\zeta$ the impact parameter of the streamline passing through the point $(r,\theta)$ :
\begin{equation}
\zeta=\frac{r\sin\theta}{2}\left[ 1+\sqrt{1+2\frac{\zeta_{\textsc{hl}}}{r}\frac{1-\cos\theta}{\sin ^2\theta}} \right]
\end{equation}
For the total specific energy density $e$, we assign a similar equation as $\rho$ but substituting the mass density by Bernoulli's invariant $e+P+\rho \Phi$ :
\begin{equation}
\begin{split}
e=\frac{1}{\gamma}  \biggl[ \bigl(e+P+&\rho\Phi\bigr)_{\infty}\frac{\zeta ^2}{r\sin\theta \left( 2\zeta-r\sin\theta \right)} \\
& - \rho \Phi + \left( \gamma - 1 \right) e_K \biggr]
\end{split}
\end{equation}
where $\gamma$ is the adiabatic index from the ideal gas equation-of-state $u=P/ (\gamma-1)$. It assures that the gravitational ballistic deflection of the initially planar flow, from infinity to the outer boundary, is taken into account. 

The downstream outer boundary condition is a continuous one and to avoid any spurious reflection of pressure waves, we set its size $r_{\text{out}}$ such as the velocities at the boundary are supersonic (see cases $\mathcal{M_{\infty}}=2$, $4$, $8$ and $16$ of Figure\,\ref{fig:density}). Typically, it requires $r_{\text{out}}\sim 8\zeta _{\textsc{hl}}$.

Concerning the inner boundary conditions, great caution must be taken. Straightforward absorbing conditions (\eg floor density and continuous velocities, provided they leave the simulation space) do alter the stability of the flow without any guarantee of fitting the continuity of the radial flows. One has to not prevent the stationary solution of \eqref{eq:eq1} and \eqref{eq:eq3} to be achieved, for example by ensuring the continuity of $\rho v_r r^2$ at the inner boundary : to achieve this, we computed the density in the inner ghost cells with a first order Taylor-Young expansion and deduced the corresponding radial velocities from the continuity of the radial mass flux $\rho v_r r^2$. For the total specific energy $e$, the conserved quantity is given by the Bernoulli's condition from \eqref{eq:eq3} considered for a permanent flow : $(e+P+\rho\varphi)v_r r^2$. The value of $v_{\theta}$ is much less critical and is also set via a simple first order expansion. Such inner boundary conditions allow the flow to reach a permanent regime as the ones described further.

\section{Results and comments}
\label{sec:results}
\subsection{Large scale structure of the shock}

\subsubsection{Spatial dimensions}

\begin{figure}
\begin{center}
\includegraphics[width=0.5\textwidth]{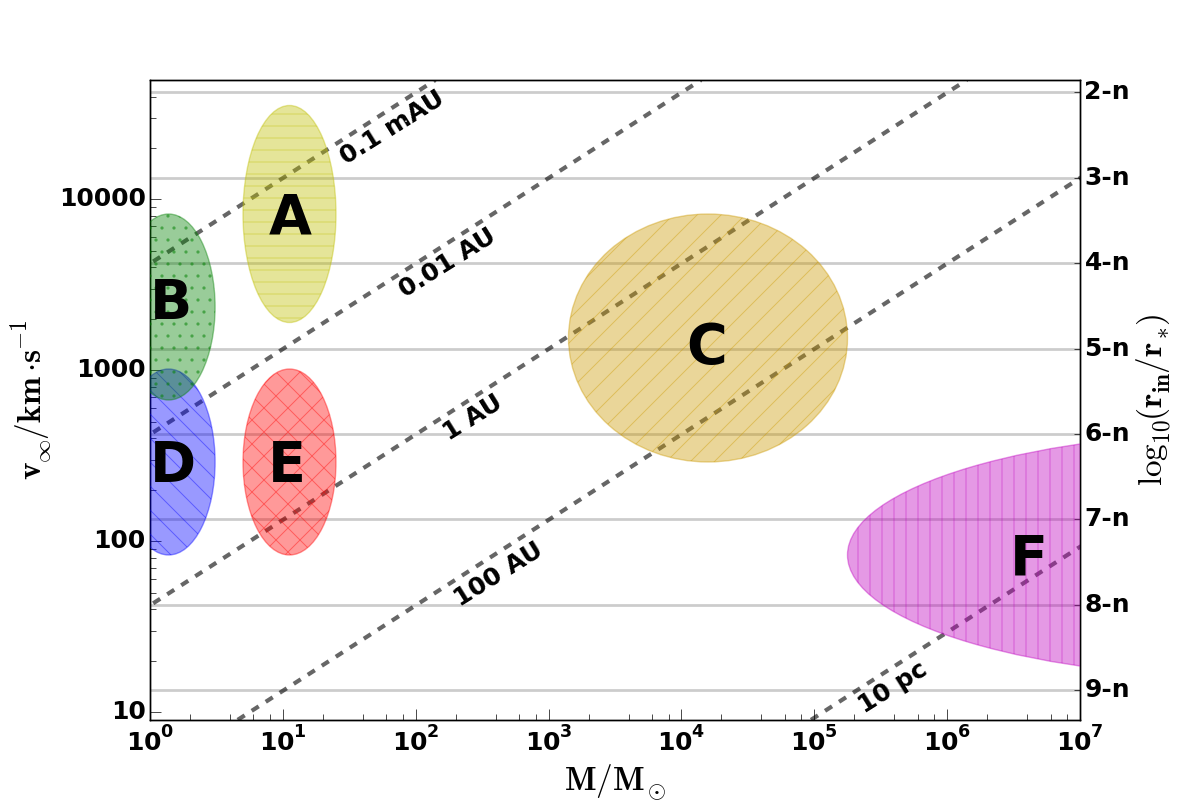}
\caption{Contour map of the ballistic extension $\zeta_{\textsc{hl}}$ of the {\sc b-h} flow as a function of the velocity at infinity and of the mass of the accreting body, confronted to an estimate of the computational cost on the right. The latter is represented by the ratio of the inner boundary radius $r_{\text{in}}$ by the Schwarzschild radius of the compact object $r_{*}$, with $n$ being given by $\zeta_{\textsc{hl}} / r_{\text{in}} = 10^n$, such as for n$=$0, the right axis indicates the physical ratio $\zeta_{\textsc{hl}} / r_{*}$. The coloured regions locate families of compact objects as detailed in the main text.}
\label{fig:dimensions_odm}
\end{center}
\end{figure}
For supersonic flows at infinity, the characteristic spatial dimension of the problem is given by the critical impact parameter $\zeta_{\textsc{hl}}$ represented on Figure\,\ref{fig:dimensions_odm}. From stellar-mass compact objects undergoing fast wind accretion ($v_{\infty}\sim 10^4$km$\cdot$s$^{-1}$) to quiescent super-massive black holes sipping the slowly drifting ambient interstellar medium, $\zeta_{\textsc{hl}}$ spans more than 10 orders of magnitude. The non-dimensionality of numerical simulations nevertheless guarantees that the ballistic behaviour upstream of the shock will not be changed from a value of $\zeta_{\textsc{hl}}$ to another ; it remains scale invariant in simulations which use $\zeta_{\textsc{hl}}$ as the length scale.

This figure also serves as a handy cheat sheet containing the main information about the spatial extension of the {\sc b-h} problem along with the computational requirements to simulate it. It also locates, as a guideline, a few families of compact objects in astrophysical systems where {\sc b-h} accretion could occur in some way :
\begin{itemize}
\item \textbf{A :} black hole high mass X-ray binaries, in particular {\sc lmc x-1} where an Onfp companion star seems to provide a particularly fast wind \citep{Orosz2008}. It could also be the case of lone runners black holes which have been accelerated in a close encounter with other black holes \citep{Sperhake2011,Lora-Clavijo2013}.
\item \textbf{B :} runaway neutron stars and {\sc s}g{\sc xb}. The first family might be illustrated by the radio pulsar PSR B2224+65 and its iconic guitar nebula \citep{Cordes1993}, albeit the pulsar wind might play an additional role in the formation of the shock compared to the simple {\sc b-h} sketch. A runaway neutron star more suitable to apply the {\sc b-h} model might be PSR J0357+3205, where the tail is too long to be accounted for by shocked pulsar wind models \citep{DeLuca2013}. For {\sc s}g{\sc xb}, it is believed that we detect the wind of an early spectral type super giant OB star accreted by a slowly-spinning neutron star orbiting on a close-in orbit \citep{Chaty2011}. Examples of such systems are Vela X-1, 4U 1907+09 \& GX 301-2.   
\item \textbf{C :} {\sc imbh}, possibly identified as Hyper Luminous X-ray sources ({\sc hlx}) whom the most honourable known member is probably ESO 243-49 HLX-1 \citep{Farrell2009}. It has been suggested, among others, that a wind accretion within a binary system might be responsible for its X-ray luminosity \citep{Miller:2014ti}. An {\sc imbh} could also be present in the Orion Nebula Cluster and undergo wind accretion from a massive stellar companion \citep{Subr2012}.
\item \textbf{D \& E :} neutron star in low mass X-ray binaries, cataclysmic variables and black holes in high mass X-ray binaries where the stellar companion wind is too faint to feed accretion, but might play an indirect role through an interplay with the Roche lobe overflowed formed accretion disc. 
\item \textbf{F :} super massive black holes accreting ambient gas of which the velocity at infinity (deduced from the kinetic energy left if it had fully escaped the super massive black hole gravitational potential) is estimated, here, from the sum of the proper motion of SgrA* and of the characteristic stellar dispersion speed \citep{Reid2003} ; it could be higher if one considers the gas launched from surrounding massive stars \citep[see][for simulations of accretion onto SgrA* along those lines]{Ruffert1994b}. Except for collisions of galaxies, the bulk motion of a super massive black hole compared to the ambient gas is likely to be negligible. 
\end{itemize}
In addition, to convert the left velocity scale to a temperature one in Kelvins for a Mach-1 flow of Hydrogen particles with $\gamma=5/3$, one can multiply 72 by the square of the velocity at infinity in km$\cdot$s$^{-1}$. Also, the reader might want to keep in mind that the {\sc ism} temperature ranges from a few 10K to 10$^6$K \citep{Ferriere2001}, the latter corresponding to a sound speed of the order of 100 km$\cdot$s$^{-1}$. The wind coming from a stellar companion could be even hotter.

To make the inner boundary as meaningful as possible, one wants $r_{\text{in}}/r_{*}$ to be as close as possible to 1 (provided we do not care about the magnetosphere influence if the central object is a neutron star), keeping in mind that each decade in space costs an additional factor of $\sim$ 40 in {\sc cpu} computational time. The ratio $\zeta_{\textsc{hl}}/r_{*}$ only depends on the wind speed at infinity (for objects whom size is given by the Schwarzschild radius) and reaches 1 for relativistic bulk motion (out of the scope of the present paper). Our simulations reach $n$ up to 4 which enables us to study accretion from reasonably fast winds (\ie a few $10^3$km$\cdot$s$^{-1}$) with\footnote{At some point where the inner boundary size reaches the order of magnitude of the compact object size or below, even without considering the magnetic effects, one can no longer neglect corrections to Newtonian mechanics.} $r_{\text{in}}\sim 10$ to $100 r_{*}$. The previous studies dealing with $r_{\text{in}} \lesssim r_{*}$ and $\zeta_{\textsc{hl}}$ in the same space simulation had to assume mildly relativistic bulk motion to make the simulation computationally affordable (with $v_{\infty}$ higher than a few $10^4$km$\cdot$s$^{-1}$) in exchange for simulations plunging so deep into the gravitational potential that a relativistic treatment is justified \citep{Font1998,Lora-Clavijo2013,Lora-Clavijo2015}. In our configuration, for n$=$4, since we do not solve the relativistic {\sc hd} equations, we have to assume that, below $10 r_{*}$ to $100 r_{*}$, matter, energy and linear momentum are trapped and do not influence the dynamics of the flow beyond, which enables us to study the systems down to the 5-n line on Figure\,\ref{fig:dimensions_odm}.

\subsubsection{Geometrical properties, profiles and stability}
\label{sec:geom}


\begin{figure*}
\begin{subfigure}{0.48\textwidth}
\begin{center}
\includegraphics[width=7.5cm, height=6.4cm]{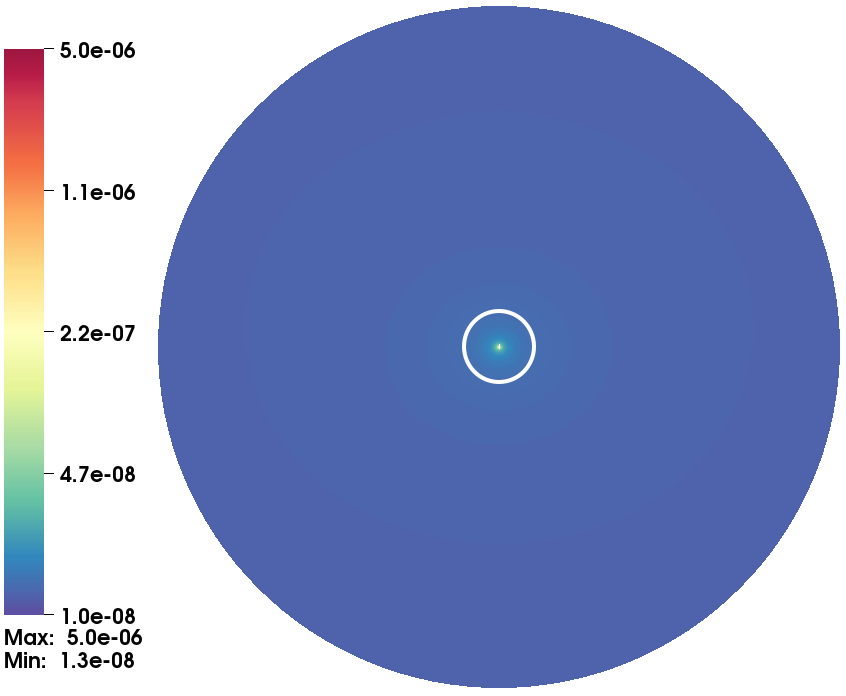} 
\label{fig:subim1}
\end{center}
\end{subfigure}
\begin{subfigure}{0.48\textwidth}
\begin{center}
\includegraphics[width=7.5cm, height=6.4cm]{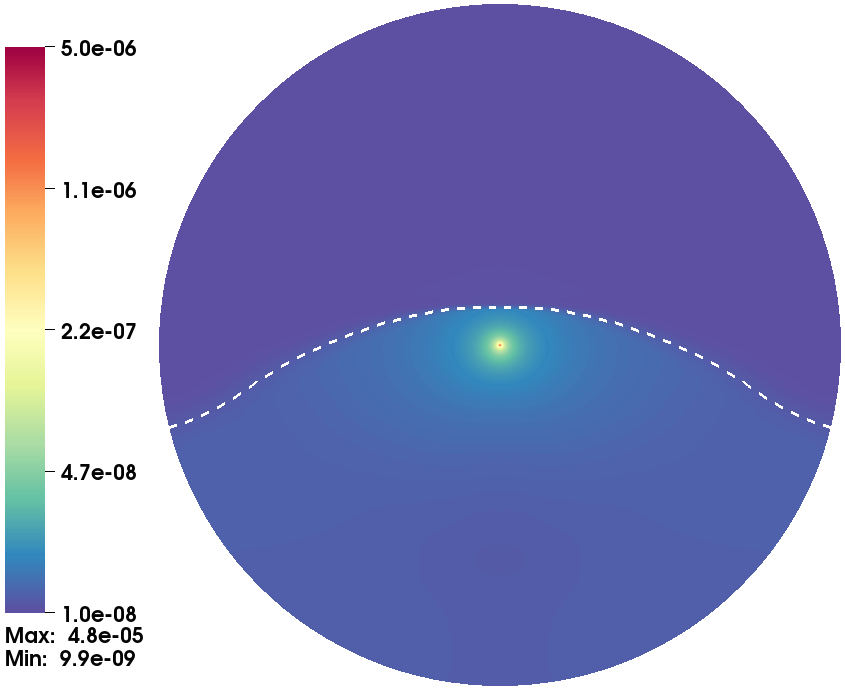}
\label{fig:subim2}
\end{center}
\end{subfigure}
\vspace*{0.8cm}\\
\hspace*{-0.2cm}
\begin{subfigure}{0.48\textwidth}
\begin{center}
\includegraphics[width=7.5cm, height=6.4cm]{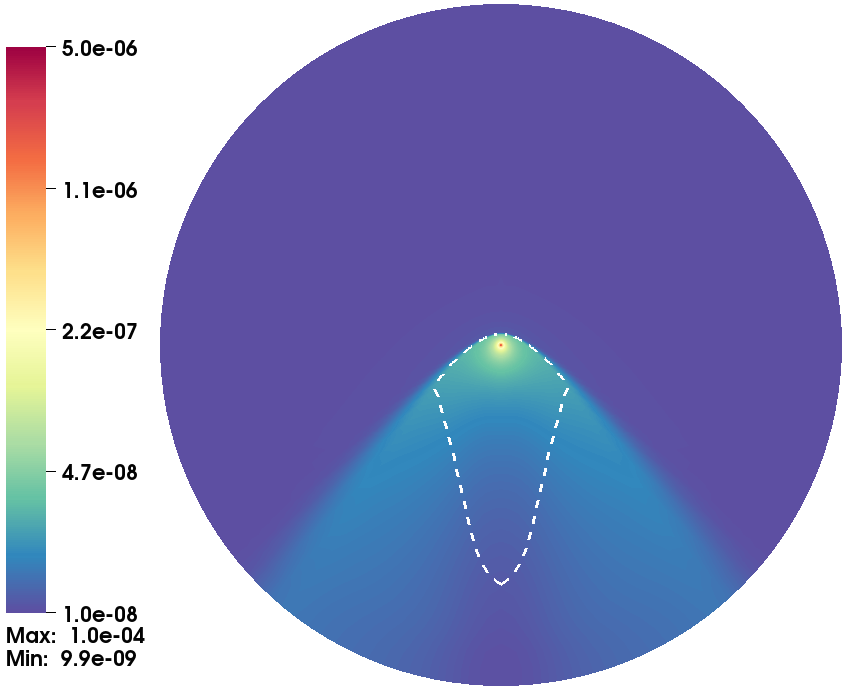}
\label{fig:subim2}
\end{center}
\end{subfigure}
\begin{subfigure}{0.48\textwidth}
\begin{center}
\includegraphics[width=7.5cm, height=6.4cm]{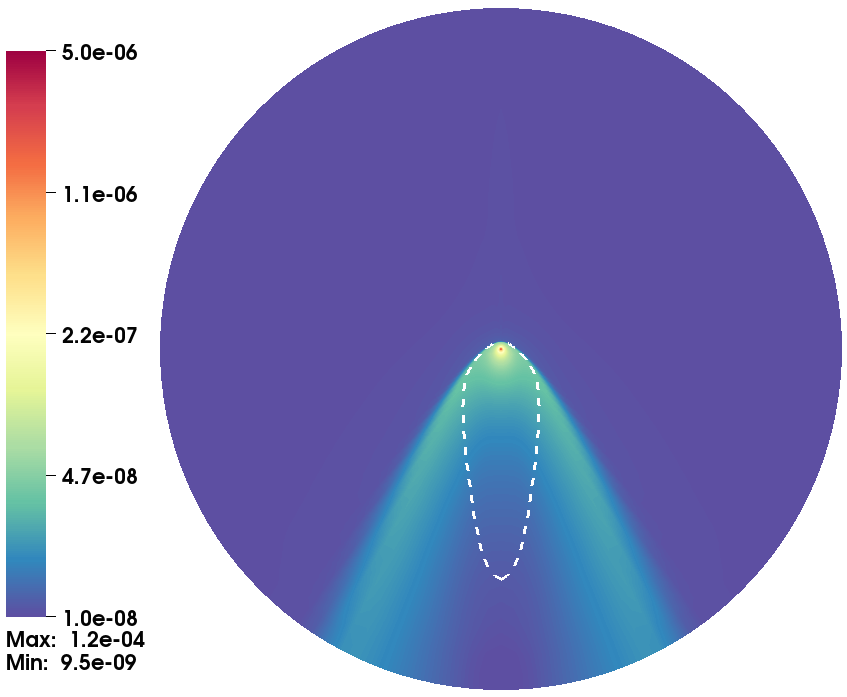} 
\label{fig:subim1}
\end{center}
\end{subfigure}
\vspace*{0.8cm}\\
\hspace*{-0.2cm}
\begin{subfigure}{0.48\textwidth}
\begin{center}
\includegraphics[width=7.5cm, height=6.4cm]{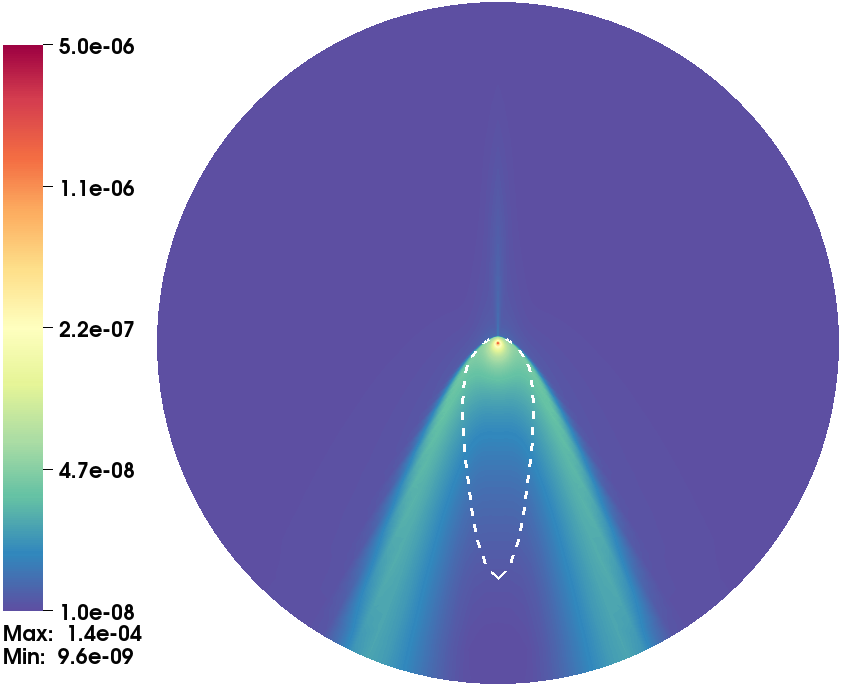}
\label{fig:subim2}
\end{center}
\end{subfigure}
\begin{subfigure}{0.48\textwidth}
\begin{center}
\includegraphics[width=7.5cm, height=6.4cm]{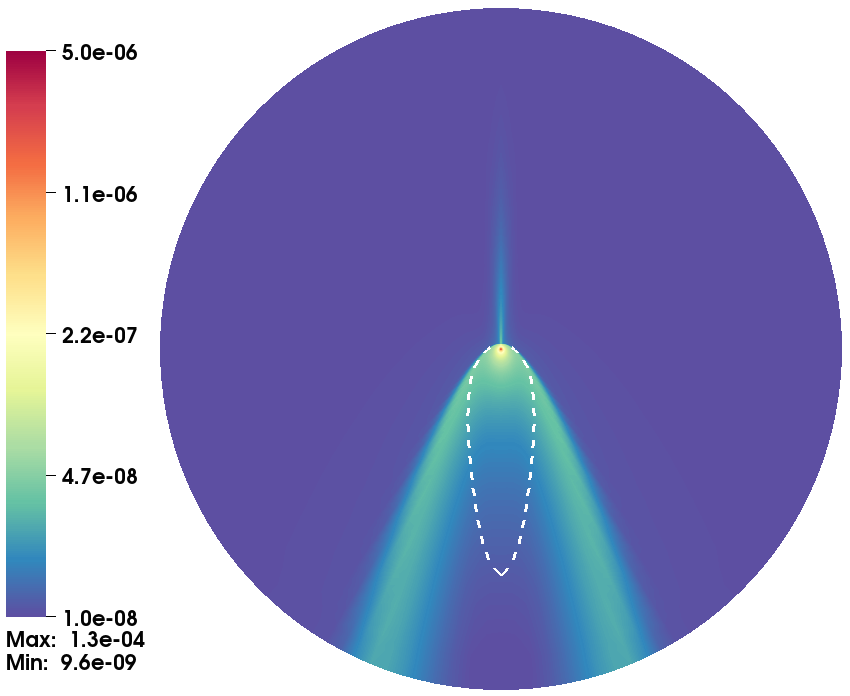}
\label{fig:subim2}
\end{center}
\end{subfigure}
\caption{Logarithmic color maps (arbitrary units) of the stationary density profiles for $r_{\text{in}}=10^{-3}\zeta _{\textsc{hl}}$ and, in reading order, $\mathcal{M_{\infty}}=0.5$, $\mathcal{M_{\infty}}=1.1$, $\mathcal{M_{\infty}}=2$, $\mathcal{M_{\infty}}=4$, $\mathcal{M_{\infty}}=8$ and $\mathcal{M_{\infty}}=16$. The outer Mach 1 surface has been plotted (not the sonic one) as a white dashed line. On the first plot, the subsonic case, the solid white line stands for the approximate size of the critical impact parameter, $\zeta_{\textsc{hl}}$ ($\zeta_{\textsc{hl}}\sim 0.15 r_{\text{out}}$).}
\label{fig:density}
\end{figure*}

\begin{figure}
\begin{center}
\includegraphics[width=0.45\textwidth]{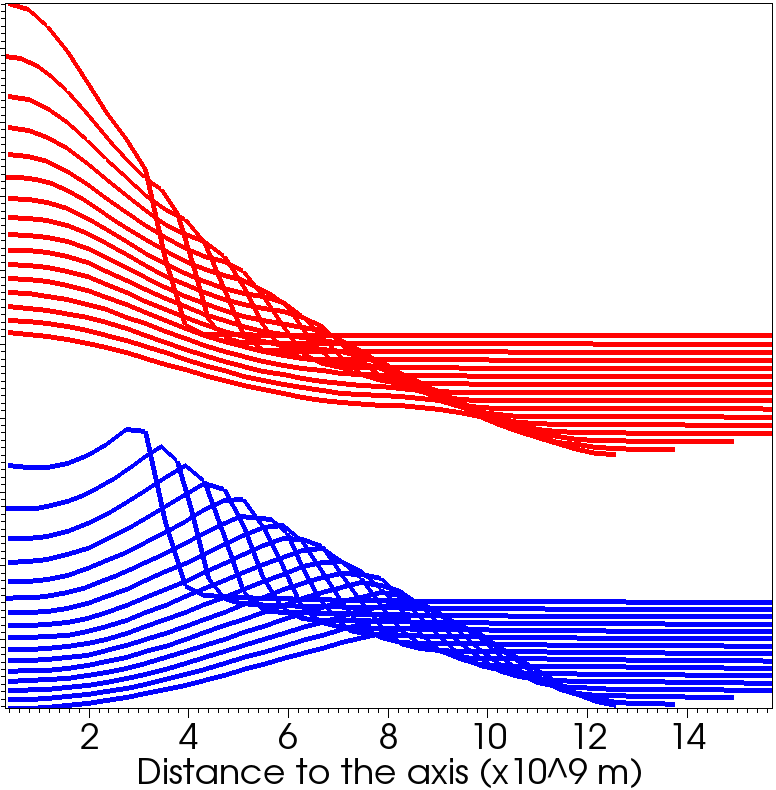}
\caption{Vertically shifted transverse profiles of density (blue) and temperature (red) along the tail of the steady-state with sampling distances along the axis ranging from $6\zeta _{\textsc{hl}}$ (lower curve) to $0.5\zeta _{\textsc{hl}}$ (upper curve). The scales are linear and the parameters are $\mathcal{M_{\infty}}=4$, $r_{\text{in}}=10^{-3}\zeta _{\textsc{hl}}$ and $\zeta _{\textsc{hl}}\sim 2.7\cdot 10^9$m.}
\label{fig:transverse_profiles}
\end{center}
\end{figure}
First comments can be made about the geometry of the shocks we obtained in the 5 supersonic configurations for which the relaxed state is represented on  Figure\,\ref{fig:density}. As long as we deal with a supersonic unperturbed flow at infinity, for any Mach number, the shock is clearly detached ; the shock front stabilises at a distance of the compact object which is large compared to the size of the inner boundary but of the order of a fraction of the critical impact parameter $\zeta_{\textsc{hl}}$ for $\mathcal{M_{\infty}}\ge 2$. The opening angle of the shock evolves along the axis, with a concave shape for the shock and smaller opening angles for larger Mach numbers.

Concerning the transverse profiles in the tail, we first notice that the shock forms a hollow cavity, in agreement with one of the solutions proposed by \cite{Bisnovatyi-Kogan1979}, the contrast between the density along the axis compared to the one at the shock front being larger for higher Mach numbers. This encouraging feature supports the idea that the bow shocks we obtained are purely the result of the gravitational beaming of the gas by the gravitational potential of the central point mass. The transverse profiles of density and temperature are shown on Figure\,\ref{fig:transverse_profiles}. The latter presents a monotonously decreasing function of the distance to the central axis, with a sharp drop at the front shock (located near the maxima of transverse density). 

\begin{figure}
\begin{center}
\includegraphics[width=0.5\textwidth]{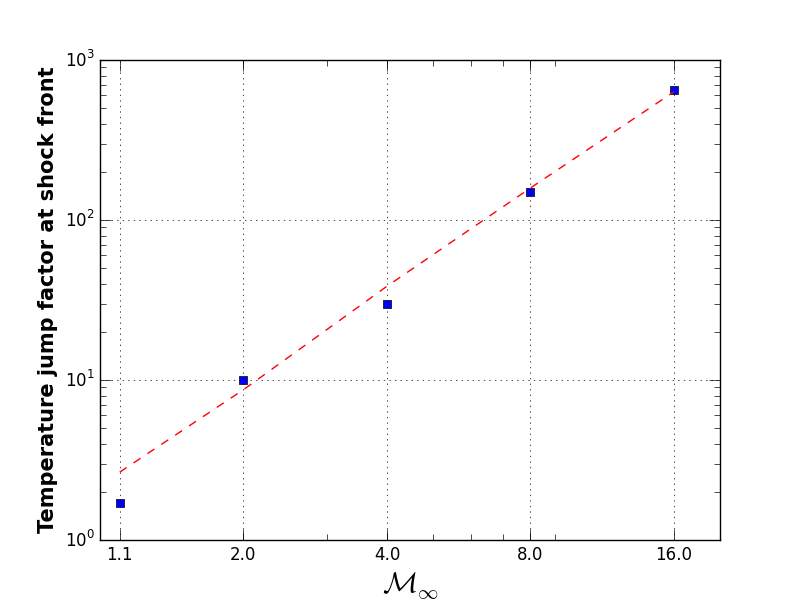}
\caption{Ratio of the upstream to the downstream temperature as a function of the Mach number at infinity. The dashed red line is the physical trend described in the text.}
\label{fig:temp_jump}
\end{center}
\end{figure}

The shock is highly non isothermal and the corresponding measured jumps in temperature are shown in blue square markers on Figure\,\ref{fig:temp_jump}. One can understand the evolution of those discontinuities with the Mach number at infinity by figuring out the corresponding Rankine-Hugoniot jump condition, with subscripts 1 and 2 corresponding respectively to upstream and downstream quantities :

\begin{equation}
\label{eq:jump}
\begin{split}
\frac{T_2}{T_1} = & \frac{\left[ \left( \gamma +1\right) + 2\gamma \left( \mathcal{M}_1^2-1\right)\right]}{\left( \gamma +1\right)^2 \mathcal{M}_{1}^2} \\
& \times \left[ \left( \gamma +1\right) + \left( \gamma -1\right) \left( \mathcal{M}_{1}^2-1\right)\right] 
\end{split}
\end{equation}
Let us now consider that the Mach number upstream $\mathcal{M}_1$ is related to the Mach number at infinity $\mathcal{M_{\infty}}$ by \eqref{eq:vr}, \eqref{eq:vt} and \eqref{eq:dens_BK} and the ballistic conversion of gravitational potential energy to kinetic. Then we get, by probing the shock at low $\theta$ angles and at a shock front $r_{\text{sh}}$ :
\begin{equation}
\label{eq:jump_T}
\mathcal{M}_1 \sim f \left[ \frac{\left( 1+f \right)^2}{4f}  \right]^{\frac{\gamma -1}{2}}\mathcal{M_{\infty}}
\end{equation}
with $f=\sqrt{1+\zeta_{\textsc{hl}}/r_{\text{sh}}}$. We used a fiducial exponential relaxation law to represent the variation of the relative position of the shock front, $\zeta_{\textsc{hl}}/r_{\text{sh}}$ between approximately $1$ and $6$ for $\mathcal{M_{\infty}}$ from $1.1$ to $16$ which gives the analytical red-dashed trend on Figure\,\ref{fig:temp_jump}. The agreement featured on this figure suggests that our numerical simulations do pass this elementary sanity check.


Although immune against non axisymmetric instabilities like the flip-flop one, our simulations could lead to longitudinal or axisymmetric transverse instabilities \citep{Cowie1977,Foglizzo2005}. However, the relative distance to the previous state - defined as the standard deviation point by point between the state values (mass density, linear momentum density, total specific energy density) at a given time step and the previous one - tends to be smaller than $0.0001$\% once the steady state is reached : no acoustic cycle takes place to maintain or amplify an oscillation in the tail, in spite of the actual production of entropy at the detached shock interface. This stability of the bow shock should not be attributed to an excessively large accretor given the ratios $\zeta_{\textsc{hl}}/r_{\text{in}}$ we used but might be linked to a solver excessively diffusive or a low angular resolution. We did run higher resolution simulations for $\mathcal{M_{\infty}}=1.1$ and $4$, with $N_{\theta}=128$ instead of $64$ (and $N_r=352$ \ie for $r_{\text{in}}/\zeta_{\textsc{hl}}=10^{-3}$). In both cases, a steady state was reached although it took longer for $\mathcal{M_{\infty}}=4$ compared to the standard resolution simulations.  

We observe a slight beaming effect along the pole due to a mesh-alignment effect. We believe it essentially alters neither the conclusions we came to nor the trends we derive for the mass accretion rate or the geometry of the shock for instance. Yet, it must be acknowledged that the quantitative information concerning the latter might give slightly underestimated positions of the forward shock because of a numerical push of artificially focused flow along the axis. As pinpointed by \cite{Blondin:2012vf}, in full 3D, such a flaw can be overcome thanks to the complementary action of two intertwined spherical meshes with orthogonal poles, the Yin-Yang mesh \citep{Kageyama2004} ; the underlying axisymmetry implied by the 2.5D mesh we use prevents us from implementing such a remedy.

\subsection{Mass accretion rates}

\subsubsection{Numerical relaxation}

\begin{figure}
\begin{center}
\includegraphics[width=0.5\textwidth]{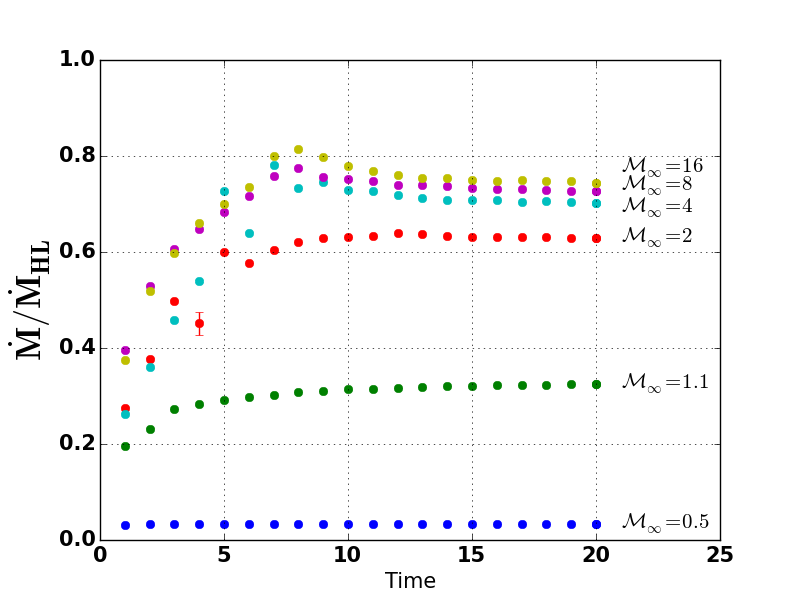}
\caption{Mass accretion rate as a function of time for different Mach numbers and $r_{\text{in}}=10^{-3}\zeta _{\textsc{hl}}$. The time unit is set by the crossing time $\zeta_{\textsc{hl}}/v_{\infty}$ and the mass accretion rate unit, by \eqref{eq:Mdot_HL}.}
\label{fig:Mdot_time}
\end{center}
\end{figure}

Figure\,\ref{fig:Mdot_time} presents the evolution of the spatially averaged mass accretion rate in the vicinity of the inner boundary, well below the shock front radius, as a function of time. It takes the system approximately 10 crossing times or less to relax the upstream outer boundary conditions specified in \eqref{sec:BC} and to reach a steady state. The normalisation of the mass accretion rates, set by \eqref{eq:Mdot_HL}, is the same for all points : $\dot{M}_{\textsc{hl}}$ does not depend on the Mach number at infinity which was changed by tuning $c_{\infty}$, not $v_{\infty}$. We do observe larger mass accretion rates for colder ambient media, with $v_{\infty}$ and $\mathcal{M_{\infty}}$ fixed, and with a saturation level which is lower than the one prescribed by the qualitative model of Hoyle \& Lyttleton (see the constant $\lambda$ in the following couple of sections).

\subsubsection{Normalisation}
\label{sec:norm}

For each Mach number, we computed a time-averaged mass accretion rate. First, we construct a non-physically motivated reference $\dot{M}_0$ whom only requirement is to verify :

\begin{equation}
\label{eq:M0_lim}
  \left\{
      \begin{array}{l}
        \dot{M}_0 \xrightarrow[\mathcal{M_{\infty}} \to \infty]{} \lambda \dot{M}_{\textsc{hl}} \\
        \dot{M}_0 \xrightarrow[\mathcal{M_{\infty}} \to 0]{} \dot{M}_B
        \end{array}
    \right.
\end{equation}
with $\lambda$ a constant further discussed in the next section. Indeed, since Bondi's derivation \eqref{eq:MBONDI} for $v_{\infty}\ll c_{\infty}$ is based on the same physical laws as the ones we implemented, the numerically computed mass accretion rates must tend towards $\dot{M}_{\textsc{b}}$ for low $\mathcal{M_{\infty}}$. On the contrary, Hoyle \& Lyttleton's qualitative approach only gives $\dot{M}_{\textsc{hl}}$ to an order of magnitude. Thus, $\dot{M}_0$ stands for a necessary asymptotic behaviour for any acceptable interpolation formula. In no case $\dot{M}_0$ should be taken as a physically meaningful formula by itself.

The structural requirement of $\dot{M}_0$ relies on a quadratic average between $v_{\infty}$ and $c_{\infty}$, in the same spirit as Bondi's interpolation formula albeit introducing a $\gamma$-dependent weighting factor for $c_{\infty}$ and a $\lambda$ one for $v_{\infty}$ :

\begin{equation}
\label{eq:vnorm}
\tilde{v}=\sqrt{\lambda v_{\infty}^2+\frac{1}{\left(\gamma -1 \right)^4}\left( \frac{2}{5-3\gamma}\right)^{\frac{5-3\gamma}{\gamma -1}}c_{\infty}^2}
\end{equation}

We introduce the modified accretion radius, $R_E$, which compares the gravitational influence of the accreting body to the joint action of kinetic and enthalpic terms at infinity :
\begin{equation}
\label{eq:RE}
R_E=\frac{GM}{\displaystyle\frac{v_{\infty}^2}{2}+\frac{c_{s,\infty}^2}{\gamma -1}}
\end{equation}
This way, the characteristic length scale from high to low Mach numbers is continuous :
\begin{equation}
\label{eq:RE_lim}
  \left\{
      \begin{array}{l}
        R_E \xrightarrow[\mathcal{M_{\infty}} \to \infty]{} \zeta_{\textsc{hl}} \\
        R_E \xrightarrow[\mathcal{M_{\infty}} \to 0]{} \frac{GM}{\displaystyle c_{s,\infty}^2/\left(\gamma -1 \right)}
        \end{array}
    \right.
\end{equation}
In the first asymptotic case, $R_E$ can be seen as the maximum impact parameter below which an amount of gravitational energy larger than the kinetic one at infinity will be converted to kinetic energy, at constant pressure and entropy (no shock). The second case is the radius below which thermodynamical properties of the gas have been significantly altered from their undisturbed state at infinity, without major change of velocities.

By writing $\dot{M}_0\propto R_{\text{E}}^2 \rho_{\infty} \tilde{v}$, we derive a formula which does not privilege a spherical nor an axisymmetric point of view. We finally make $\dot{M}_0$ verify the empirical condition \eqref{eq:M0_lim} by setting the proportionality constant to $\pi$ :
\begin{equation}
\label{eq:norm}
\dot{M}_0=\pi R_{\text{E}}^2 \rho_{\infty} \tilde{v}
\end{equation}
With $\dot{M}_0$, we have a homogeneous normalisation variable to compare the mass accretion rates to at any $\mathcal{M_{\infty}}$.

\subsubsection{Permanent mass accretion rates}
\label{sec:mdot}
\begin{figure}
\begin{center}
\includegraphics[width=0.5\textwidth]{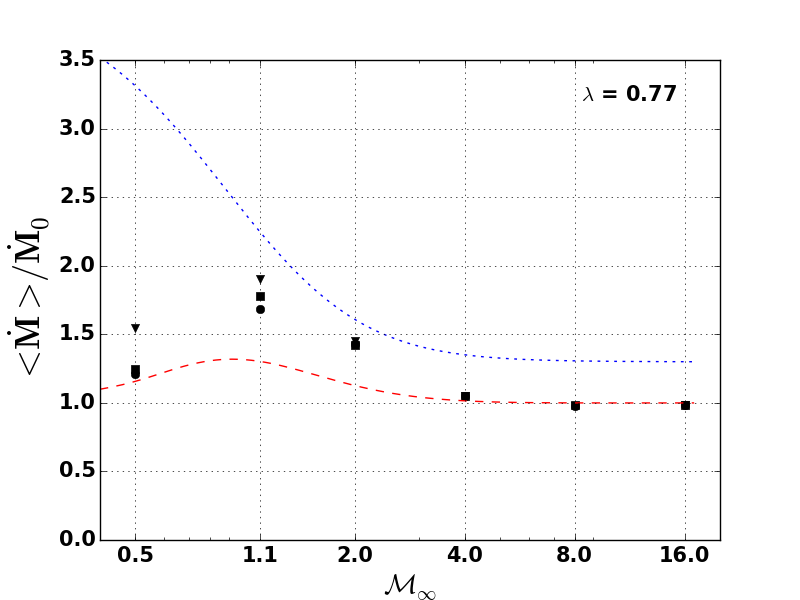}
\caption{Mass accretion rates $\dot{M}$ as a function of the Mach number $\mathcal{M_{\infty}}$, normalized with the empirical mass accretion rate given by \eqref{eq:norm}. In black are the numerically computed ones for $\zeta_{\textsc{hl}} / r_{\text{in}} = 10^2$ (triangles), $\zeta_{\textsc{hl}} / r_{\text{in}} = 10^3$ (squares) and $\zeta_{\textsc{hl}} / r_{\text{in}} = 10^4$ (circles, only computed for the three lowest Mach numbers). The blue dotted line is the Bondi interpolation formula and the red dashed is the FR96's one.}
\label{fig:Mdot_Mach}
\end{center}
\end{figure}

The steady-state mass accretion rates $\langle\dot{M}\rangle$ are computed by averaging the instantaneous ones, and the error bars from the dispersion, both for $t>10$ crossing times, once the steady state is reached. The obtained $\langle\dot{M}\rangle$ are presented on Figure\,\ref{fig:Mdot_Mach}. The error bars are smaller than the markers' size but for small $\mathcal{M_{\infty}}$, the data dispersion is dominated by systematics, the influence of which is given by the two sets of points from the different inner boundary sizes.
As expected, the traditional Bondi formula, in blue dotted on Figure\,\ref{fig:Mdot_Mach}, matches the Hoyle-Lyttleton mass accretion rate at high Mach numbers, which itself turns out to be an overestimation of the actual mass accretion rate by approximately 30\% : it is consistent with the previous reports of $\dot{M}_{\textsc{hl}}$ being a slight overestimation of the numerically observed mass accretion rates in the asymptotically supersonic regime - see, \eg, Figure\,7 of \cite{Edgar:2004ip}. Yet, since \eqref{eq:interpolBH} does not properly include thermodynamics (\eg it does not depend on the adiabatic index), it is a poor estimation in the subsonic regimes, overestimating $\dot{M}$ by a factor up to 4 for $\gamma=5/3$ - which corresponds to the analytically derived limit of $\dot{M}_{\textsc{bh}}/\dot{M}_{\textsc{b}}$ at low Mach numbers given in the consistent and comprehensive review of \cite{Ruffert1994a}. But surprisingly enough, $\dot{M}_{\textsc{bh}}$ is not too far off in the mildly supersonic case, where the Hoyle-Lyttleton formula is invoked with the velocity at infinity replaced by its quadratic average with the sound speed at infinity.

\begin{figure}
\begin{center}
\includegraphics[width=0.45\textwidth]{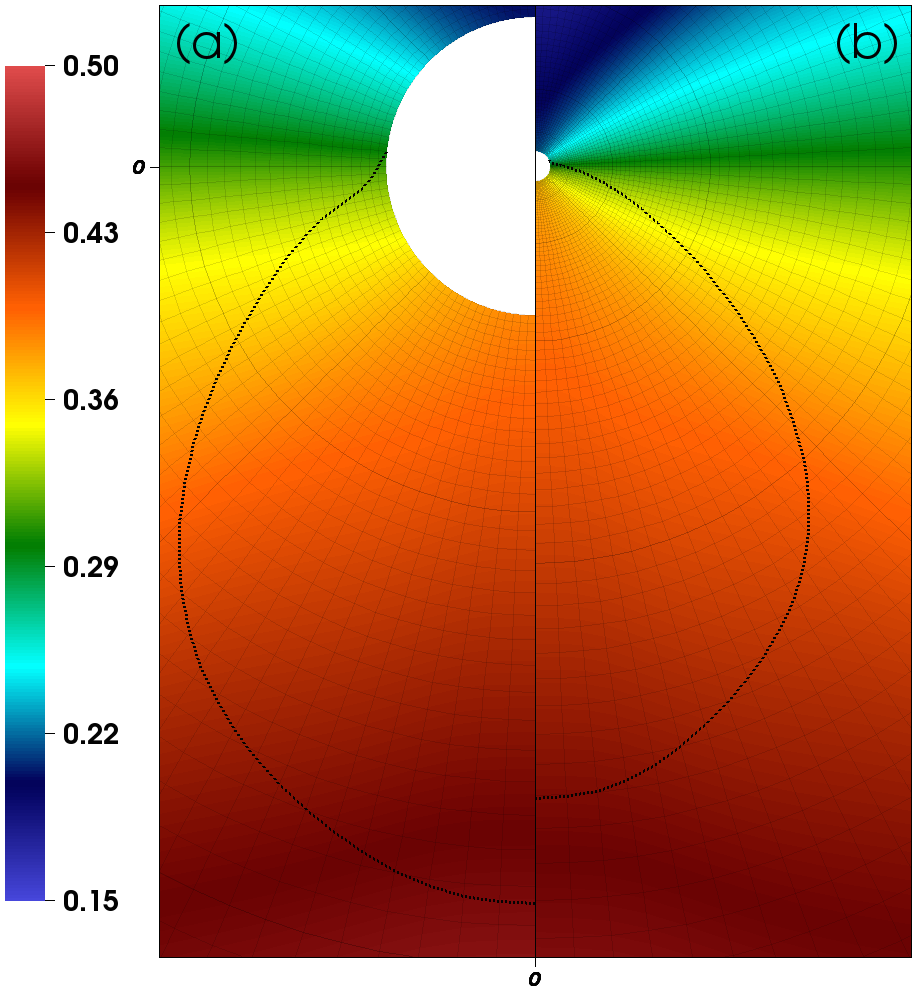}
\caption{Zoom in by a factor 500 on the central area of the simulation. Colour map shows the local mass accretion rate $\rho v_r r^2$, with $\mathcal{M_{\infty}}=2$, for (a) $r_{\text{in}}=10^{-2}\zeta _{\textsc{hl}}$ and (b) $r_{\text{in}}=10^{-3}\zeta _{\textsc{hl}}$. The scale is linear (arbitrary units) and the inflowing mass is counted positively. The thick dotted black line is the sonic surface, anchored to the inner boundary in both cases. The logarithmic mesh is shown by thin radial and azimuthal solid black lines.}
\label{fig:diff_rin}
\end{center}
\end{figure}

Foglizzo and Ruffert's interpolation formula, represented in red dashes on Figure\,\ref{fig:Mdot_Mach}, is much more physically motivated. They split the mass flux in 3 different components (but only two of them are defined in a detached shock configuration as ours) : the mass flowing through the sonic surface and the mass flux through the angular sector where the flow is still subsonic (see Figure\,\ref{fig:diff_rin} for a zoom in on the vicinity of the inner boundary). Neglecting the latter, they derived an interpolation formula to determine the former :

\begin{equation}
\label{eq:interpolFR}
\begin{split}
\dot{M}_{\textsc{fr}}=\dot{M}_\textsc{b} &\left[ \frac{\left(\gamma +1\right)\mathcal{M_{\text{eff}}}^2}{2+\left(\gamma -1\right)\mathcal{M_{\text{eff}}}^2} \right]^{\frac{\gamma}{\gamma -1}} \left[ \frac{\gamma +1}{2\gamma \mathcal{M_{\text{eff}}}^2-\gamma +1}\right]^{\frac{1}{\gamma -1}} \\
& \times \left[ 1+\frac{\gamma -1}{2} \mathcal{M_{\infty}}^2 \right]^{\frac{5-3\gamma}{2\left(\gamma -1\right)}}
\end{split}
\end{equation}
with the second line being equal to 1 for $\gamma=5/3$ and with $\mathcal{M_{\text{eff}}}$ the effective Mach number which depends on a free parameter $\lambda$ assessing the aforementioned discrepancy between $\dot{M}_{\textsc{hl}}$ and the observed mass accretion rate at high $\mathcal{M_{\infty}}$ :
\begin{equation}
\label{eq:lambda}
\begin{split}
\dot{M}_{\textsc{fr}}\xrightarrow[\mathcal{M_{\infty}} \to \infty]{}\lambda \dot{M}_{\textsc{hl}}
\end{split}
\end{equation}
\begin{equation}
\label{eq:Meff}
\frac{\mathcal{M_{\text{eff}}}}{\mathcal{M_{\infty}}}\sim \frac{1}{2^{\gamma}\lambda ^{\frac{\gamma -1}{2}}} \sqrt{\frac{2}{\gamma}} \frac{\left( \gamma +1 \right)^{\frac{\gamma +1}{2}}}{ \left( \gamma -1 \right)^{\frac{5\left(\gamma -1\right)}{4}} \left( 5-3\gamma \right)^{\frac{5-3\gamma}{4}} }
\end{equation}
Physically, $\lambda$ accounts for an essential non-ballistic feature of the flow, whatever high the Mach number is. As a consequence, Foglizzo and Ruffert's formula must be understood as a lower limit since it neglects the matter being accreted from a subsonic region the angular extension of which becomes larger as $\mathcal{M_{\infty}}$ decreases.

Although pretty similar to Ruffert's numerical conclusions \citep{Ruffert1994c,Ruffert1994}, our results do not show any decreasing trend at high Mach number and match the interpolation formula $\dot{M}_{\textsc{fr}}$ in the asymptotically supersonic regime for $\lambda\sim 0.77$, down to $\mathcal{M_{\infty}}=4$, at a few percents precision level. However, the precision on $\lambda$ could be limited by the numerical beaming along the $\theta=0$ axis\footnote{Visible on Figure\,\ref{fig:density} for $\mathcal{M_{\infty}}\ge 8$.} mentioned in the section \ref{sec:geom}, although compensated by a similar depletion along the $\theta=\pi$ axis. Concerning the resolution dependency of our simulations, with the higher resolution simulations described in \ref{sec:geom}, we did not observe any significant change of the mass accretion rate beyond a 2\% level in the case $\mathcal{M_{\infty}}=4$. Yet, it turned out that the $\mathcal{M_{\infty}}=1.1$ case gave accretion rates 5 to 10\% lower in the high resolution runs, which emphasizes the larger systematic uncertainties for mildly supersonic flows. 

The main feature of the axisymmetric {\sc b-h} accretion flow, its amplification by a few 10\% of $\dot{M}$ around $\mathcal{M_{\infty}}=1$ compared to the interpolation formulae \eqref{eq:interpolFR} and \eqref{eq:norm} verifying $\dot{M}\xrightarrow[]{\mathcal{M_{\infty}} \to 0}\dot{M}_{\textsc{b}}$ and $\dot{M}\xrightarrow[]{\mathcal{M_{\infty}} \to \infty}\sim\dot{M}_{\textsc{hl}}$, is also contained in our results. It must be noticed that for $\mathcal{M_{\infty}} = 0.5$, $1.1$ \& $2$, our mass accretion rates converge towards FR96's interpolation formula as $r_{\text{in}}/\zeta_{\textsc{hl}}$ drops, without fully settling down (this in agreement with the zero value of the sonic radius). Since the sonic surface is anchored to the accretor, there are always directions of accretion where the flow is not supersonic. The smaller the inner boundary size, the closer the simulation is to the model drawn in FR96. It can also be seen from direct visualization of the sonic surface. It tends to occupy a larger angular region around the inner boundary for smaller inner boundary radii. Given those elements, those numerical $\dot{M}$ must be seen as upper limits of the ones one would get for a smaller absorbing\footnote{In the meaning of the inner boundary conditions described in \ref{sec:BC}.} inner boundary. \\

\subsection{The sonic surface}
\label{sec:sonic_surface}
\begin{figure}
\begin{center}
\includegraphics[width=0.45\textwidth]{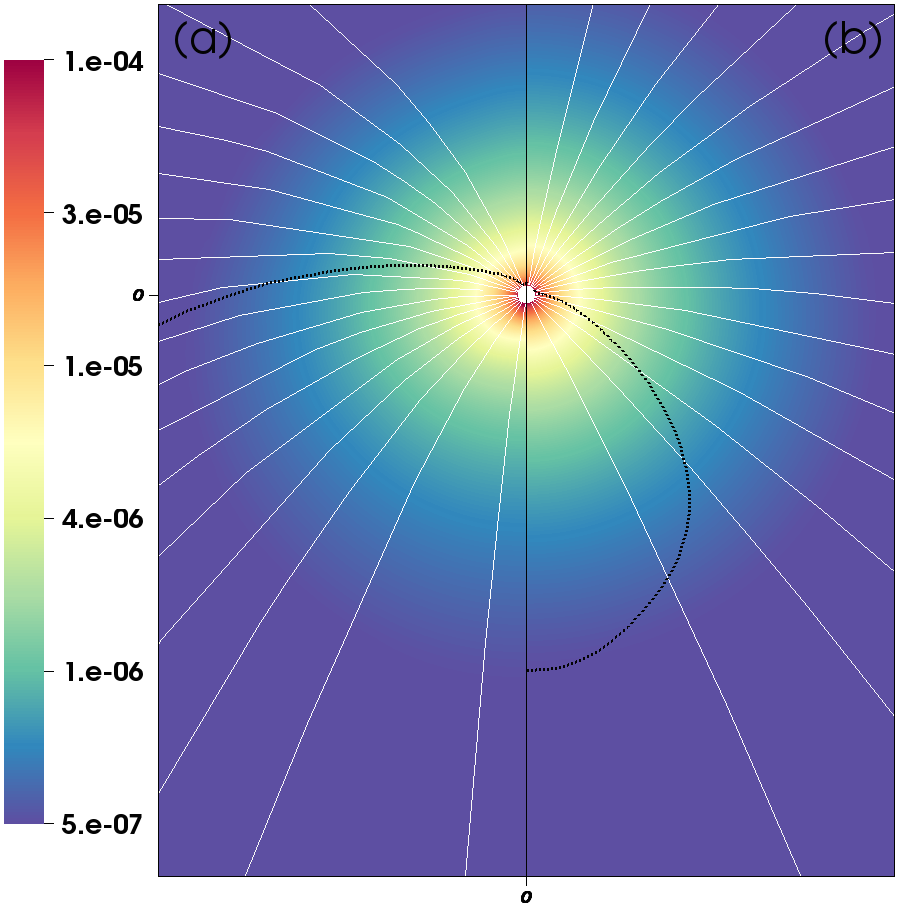}
\caption{Logarithmic color map of the density (arbitrary units), with $r_{\text{in}}=10^{-3}\zeta _{\textsc{hl}}$, for (a) $\mathcal{M_{\infty}}=8$ and (b) $\mathcal{M_{\infty}}=2$. The white lines are the streamlines and the thick dotted black line is the sonic surface, anchored to the inner boundary in both cases.  This is a zoom in by a factor 300 on the central area of the simulation.}
\label{fig:diff_Mach}
\end{center}
\end{figure}
An important conclusion of those simulations is the confirmation of Foglizzo and Ruffert's analytical prediction about the topology of the sonic surface for an adiabatic flow with $\gamma=5/3$ : whatever the size of the inner boundary (Figure\,\ref{fig:diff_rin}) or the Mach number of the supersonic flow (Figure\,\ref{fig:diff_Mach}), the sonic surface is always anchored to the inner boundary and it extends along the wake of the accretor. For supersonic flows, the density distribution is mostly isotropic and the streamlines, radial, in the vicinity of the inner boundary. Thus, the accretion is regular in the sense that there is no infinite mass accretion rate direction, although the local mass accretion rate is enhanced by a factor of a few units in the back hemisphere compared to the front hemisphere. As a consequence, the non isotropy of the mass accretion rates around the inner boundary is mostly due to the non isotropy of the velocity field ; along the virtual accretion line, in the wake of the accretor, the flow has been reaccelerated more than upstream behind the shock.





\section{Summary \& outlook}

We designed a numerical setup optimized to embrace the whole dynamics of a 2.5D planar hydrodynamical Bondi-Hoyle accretion flow onto a body whom size is 3 to 5 orders of magnitude smaller than its accretion radius, reproducing typical weakly or non-magnetized compact objects. Despite the challenging dynamics, we are able to characterise the mass accretion rates for Mach numbers at infinity ranging from 0.5 to 16, along with the geometrical properties of the flow : the bow shock forming for supersonic flows, the internal depletion of mass along the accretion line (beyond the stagnation point) and the temperature jump at the shock are all in agreement with previous analytic solutions \citep{Bisnovatyi-Kogan1979}. Those simulations also emphasized the relevance of the sonic surface in the derivation of the mass accretion rates. As first suggested by FR96, anchoring of the latter to the accretor surface allows one to set an analytical lower limit on the mass accretion rates and to understand, at least qualitatively, its dependence on the Mach number for mildly supersonic flows. We evaluated the underestimation factor induced by such an approach by accounting for the whole mass flow, not only through the supersonic surface but also in the subsonic angular section. It is worth mentioning that the anchoring of the sonic surface, shown here for the first time, is a strong consistency check which brings robustness to the physical accuracy of those simulations. Furthermore, we observed the relaxation towards a numerically stationary regime, which does not present any significant oscillatory behaviour, discarding the growth of axisymmetric instabilities with this level of numerical dissipation. Concerning non axisymmetric instabilities like the so-called flip-flop instability, unfolding those 2.5D setups in full 3D would enable us to investigate whether its amplitude is similar to the 2D cases. 

The use of a proper energy equation alleviated the introduction of a polytropic index and ensured the anchoring of the sonic surface for any monoatomic ideal gas ; it also naturally handled the jump conditions at the shock. Yet, we relied on an adiabatic assumption which might overestimate the gas heating as matter flows in. Besides, we neglect radiative feedbacks which are likely to play a role through, for instance, the radiative pressure. Indeed, comparing the Eddington luminosity to the Hoyle-Lyttleton mass accretion rate with a fiducial efficiency of 10\% gives :

\begin{equation}
\begin{split}
\frac{10\% \dot{M}_{\textsc{hl}}c^2}{L_{\text{Edd}}} \sim 4\% \left( \frac{M}{20M_{\odot}} \right) &\left( \frac{v_{\infty}}{10^6\text{m}\cdot\text{s}^{-1}} \right)^{-3} \\
& \left( \frac{\rho _{\infty}}{10^{-11} \text{kg}\cdot \text{m}^{-3}} \right) 
\end{split}
\end{equation} 

Taking into account this radiative term in the energy equation is necessary for slower winds, {\sc imbh} \citep{Park2013} or {\sc smbh} \citep{Novak2011}. Persistent radiative feedback of accreting stellar-mass black holes on their host primordial galaxies is also under current investigation, along with the central supermassive black hole influence \citep{Wheeler2011}. Concerning orbital effects in binaries \citep[see][for {\sc sph} simulations]{Theuns1993,Theuns1996}, the present study has not included them but serves as a test case for a forthcoming work accounting for their effects. Yet, if the orbitally induced torque remains small enough, those axisymmetric simulations are not expected to depart much from the actual configuration of {\sc s}g{\sc xb} where the mass transfer is believed to occur mainly through fast stellar winds. Thanks to our numerical setup, which reconciles the requirement for physical size of the accretor together with the necessity to include the accretion radius within the simulation space, we stepped through a numerical threshold beyond which we expect the wind accretion process at stake in binaries to be simulated comprehensively.

%
%

\section*{Acknowledgments}

We thank the referee for an encouraging and incisive report which improved the quality of the present paper. We gratefully thank Thierry Foglizzo for fruitful exchanges and the {\sc mpi-amrvac} development team, in particular Oliver Porth, Rony Keppens and Zakaria Meliani, for useful assistance. We acknowledge the financial support from the UnivEarthS Labex program of Sorbonne Paris Cit\'e (ANR-10-LABX-0023 and ANR-11-IDEX-0005-02). IEM acknowledges financial support from the {\sc ens} of Cachan through a \textit{contrat doctoral sp\'ecifique pour normaliens} and from the University Paris 7 Diderot. The numerical simulations were ran on the Fran\c cois Arago Centre facilities in Paris 7 and using the HPC resources from GENCI-CINES (Grant c2015047469).

\newpage


\begin{normalsize}
\bibliographystyle{agsm}
\bibliography{/Users/ielm/Work/LateX/MNRAS_templates/bhl_25D/Refereed/To_send_to_MNRAS_revised_version/Source_files/article_bhl_25D_no_url.bib}
\end{normalsize}

\bsp

\label{lastpage}

\end{document}